\documentclass[aps,prd,twocolumn,showpacs,preprintnumbers,nofootinbib,
               superscriptaddress]{revtex4-1}
\usepackage{graphicx, fancybox,bm}
\usepackage{amsmath,amssymb}
\usepackage{color}

\newcommand{\e}{\mbox{e}}

\begin{document}

\preprint{BI-TP 2012/10}

\title{
Thermal mass and dispersion relations of quarks 
in the deconfined phase of \\
quenched QCD
}

\author{Olaf Kaczmarek}
\email{okacz@physik.uni-bielefeld.de}
\affiliation{
Fakult\"at f\"ur Physik, Universit\"at Bielefeld, D-33615 Bielefeld, Germany}

\author{Frithjof Karsch}
\email{karsch@bnl.gov}
\affiliation{
Fakult\"at f\"ur Physik, Universit\"at Bielefeld, D-33615 Bielefeld, Germany}
\affiliation{
Brookhaven National Laboratory, Bldg. 510A, Upton, NY 11973, USA
}

\author{Masakiyo Kitazawa}
\email{kitazawa@phys.sci.osaka-u.ac.jp}
\affiliation{
Department of Physics, Osaka University, Toyonaka, Osaka, 560-0043, Japan
}

\author{Wolfgang S\"oldner}
\email{wolfgang.soeldner@physik.uni-regensburg.de}
\affiliation{
Institut f\"ur Theoretische Physik, Universit\"at Regensburg, D-93040 Regensburg, Germany}

\begin{abstract}

Temporal quark correlation functions are analyzed in quenched 
lattice QCD for two values of temperature above the critical
temperature ($T_c$) for deconfinement,
$T=1.5T_c$ and $3T_c$. A two-pole ansatz for the quark
spectral function is used to determine the bare quark
mass and the momentum dependence of
excitation spectra on large lattices of size up to
$128^3\times16$. 
The dependence of the quark correlator on these parameters
as well as the finite volume dependence of the
excitation energies are 
analyzed in detail in order to examine the reliability of 
our analysis.
Our results suggest the existence of quasi-particle peaks 
in the quark spectrum. We furthermore find evidence that the 
dispersion relation of the plasmino mode has a minimum at non-zero 
momentum even in the non-perturbative region near $T_c$.
We also elaborate on the enhancement of the quark correlator
near the chiral limit which is observed at $T=1.5T_c$ on about 
half of the gauge configurations. We attribute this to the 
presence of near zero-modes of the fermion matrix that are 
associated with non-trivial topology of the gauge configurations.

\end{abstract}

\date{\today}

\maketitle


\section{Introduction}
\label{sec:Intro}

At asymptotically high temperature ($T$) strongly interacting
matter, described by Quantum Chromodynamics (QCD), behaves like 
an almost free gas of quarks and gluons.
Excitations of the medium in this temperature range are well 
described by dispersion relations of free quarks and gluons.
As  $T$ is lowered, the temperature-dependent running 
coupling constant, $g$, increases and the excitation 
spectra of the elementary quark and gluon fields are modified.
Eventually they deviate substantially from those of free particles.
In particular at those low temperatures close but above the
QCD phase transition, $T_c$, at which experiments at the
Relativistic Heavy Ion Collider (RHIC) \cite{RHIC} found 
evidence for the existence of strong correlations in
hot and dense matter, it is of interest to understand
the fate of the elementary excitations, their quasi-particle
properties and the structure of dispersion relations. 

>From leading-order resummed perturbation theory it is 
known that collective excitations of the quark and gluon fields 
acquire mass gaps (thermal masses) and decay rates proportional 
to $gT$ and $g^2T$, respectively \cite{LeBellac}.
Since at leading order the decay rates in units of $T$ 
parametrically grow faster than the thermal masses 
as $T$ is lowered, it is na\"ively expected 
that quasi-particle modes of these fields cease to exist at low 
temperatures at which $g$ is large, even in the deconfined phase.
On the other hand, the experimentally observed quark number scaling 
of elliptic flow suggests the existence of quasi-particles 
having quark quantum numbers at the early stage of fireballs 
created in heavy ion collisions \cite{Fries:2003kq}.
Also the behavior of quark number susceptibilities and 
higher-order cumulants of quark number fluctuations,
calculated in lattice QCD, suggests the existence of 
quasi-particles even near $T_c$ \cite{fluctuations,newfluct}.

One of the striking features of the quasi-particle excitations 
of quarks, found in leading order perturbation theory, is that the 
in-medium quark dispersion relation splits into two branches, the 
normal and plasmino modes \cite{plasmino}. Moreover, the latter has 
a non-trivial minimum at non-vanishing momentum of order $gT$ 
\cite{plasmino}.
Such a spectrum of fermions is, in fact, quite common.
In the weak coupling and high $T$ limit it is realized
not only for QCD,
but also in a wide class of models where 
the fermion couples to a boson. Away from this limiting situation
the fermion spectrum has 
been investigated in various theoretical settings 
\cite{BBS92,KKN06,Kitazawa:2007ep,Harada:2008vk,
Muller:2010am,Mueller:2010ah,Qin:2010pc,Satow:2010ia,
Nakkagawa:2011ci,Hidaka:2011rz}.
An interesting observation made in these studies is 
that the fermion spectrum exhibits a variety of structures 
depending on parameters of the system.
For example, it has been pointed out that the fermion 
spectrum has a multi-peak structure when the mass of bosons 
that couple to the fermion is comparable with $T$ \cite{KKN06}:
In addition to the normal and plasmino peaks having 
thermal masses, there appears an additional, third dispersion 
which crosses the origin of the energy-momentum plane.
The existence of such a multi-peak structure is also 
suggested in Refs.~\cite{Harada:2008vk,Qin:2010pc,
Hidaka:2011rz}.
It is thus interesting to investigate the fate of this 
third branch in the quark spectrum in the 
non-perturbative region near but above $T_c$.

In order to gain insight into the quasi-particle nature
of quarks in the non-perturbative region, analyses of 
the spectral properties of quarks have recently been carried 
out on the lattice QCD \cite{KK07,KK09,Hamada:2010zz}.
Lattice calculations, performed in the quenched
approximation of QCD in Landau gauge \cite{KK07,KK09} for 
$1.25 < T/T_c < 3.0$,
indicated that a plasmino mode with a distinctively
different dispersion relation contributes to the
quark spectral function in addition to the normal
mode even at temperatures close to $T_c$.
In the chiral limit these modes have identical quasi-particle 
masses that are, in the range of temperature analyzed, 
approximately proportional to $T$. 
It is notable that similar results have been reported in 
the analysis of the quark spectrum with Schwinger-Dyson 
equations \cite{Mueller:2010ah,Qin:2010pc}.

The lattice QCD calculations presented in Refs.~\cite{KK07,KK09},
however, showed that the quasi-particle masses
and dispersion relations of quarks are quite
sensitive to finite volume ($V$) effects. 
The quark thermal mass drops as the aspect ratio
$V^{1/3}T\equiv N_\sigma/N_\tau$ increases, 
and it does not converge at the value on the largest lattice 
in Refs.~\cite{KK07,KK09} with $N_\sigma/N_\tau=4$, 
while the lattice spacing $a$ used in these studies 
was found to be fine enough to suppress lattice 
discretization effects.
The origin of such a strong $V$ dependence is well 
understood in terms of discretization effects
in momentum space.
For $N_\sigma/N_\tau=4$ the lowest non-zero momentum 
is $p_{\rm min} = 2\pi (N_\tau/N_\sigma)T \simeq 1.6T$ 
for periodic boundary condition along the spatial 
direction, which is still significantly larger than $T$.
On lattices with such an aspect ratio effects of 
thermally-excited particles with momenta of order $T$,
which are responsible for the emergence of the thermal 
mass in terms of perturbation theory \cite{LeBellac,HTL}, 
will thus not be properly incorporated.

In the present study, 
we analyze the quark correlation function on lattices 
with much larger aspect ratio, $N_\sigma/N_\tau=8$.
With this aspect ratio, the lowest non-zero momentum, 
$p_{\rm min} \simeq 0.79T$, is lower than $T$, and effects of low 
momentum modes are better incorporated in numerical calculations.
To extract the spectral function from Euclidean lattice 
correlators, we use the two-pole ansatz 
as in the previous studies \cite{KK07,KK09}.
We show that this ansatz reproduces the lattice correlator 
well on the largest lattice over rather wide ranges 
of bare quark masses and momenta. We use the new large
volume results to improve the extrapolation of the 
thermal quark mass values to the infinite volume limit.
We obtain a value which is about $10\%$ smaller 
in magnitude than the one obtained in the earlier analysis.
The large spatial volume also allows one to directly analyze 
the momentum dependence of excitation spectra in more detail.
We show that the dispersion relation for the plasmino mode
obtained with the two-pole ansatz has a minimum at non-zero
momentum.

To clarify physical consequences for the quark spectral function 
that can be extracted from the analysis with the two-pole 
ansatz, we also take a closer look at the Euclidean quark 
correlator on the lattice. 
We show that with the present statistics the two-pole 
ansatz can give small $\chi^2/{\rm dof}$ even if the quark 
spectral function does not consist solely of two peaks.
Nevertheless, we argue that the analysis performed by us
for finite ranges 
of parameters indicates the existence of quasi-particle 
peaks corresponding to the normal and plasmino modes having 
thermal masses near but above $T_c$.

At $T=1.5T_c$, we find that on about half
of the gauge configurations the quark correlator shows an 
enhancement deviating from the generic behavior near the 
chiral limit. 
We show that this behavior is sensitive to
the chirality of the quark propagator and argue
that it is expected to arise from gauge configurations
with non-trivial topology which are still abundant at
this value of the temperature. We do not find any of 
these effects at the larger temperature $T=3 T_c$.

This paper is organized as follows.
In the next section we summarize the basic properties 
of the quark correlator and present our simulation setup. 
In Sec.~\ref{sec:topology}, we discuss the enhancement of 
the quark correlator near the chiral limit observed 
on some gauge configurations for $T=1.5T_c$. 
In Sec.~\ref{sec:spectra} we analyze the dependences of 
the quark spectral function on the bare quark mass and momentum.
The last section is devoted to a short summary.
In appendix~\ref{sec:S-rho} we discuss the relation between
spectral function and Euclidean correlator to clarify the
sensitivity of the lattice correlator on the quark spectrum.

\section{Quark spectral function}
\label{sec:setup}

\begin{table}
\begin{center}
\begin{tabular}{cccccccc}
\hline
\hline
$T/T_c$ & $\beta$ & $N_\sigma$ & $N_\tau$ & $\kappa_c$ & $N_{\rm conf}$ & $N_{\rm top}$ \\
\hline
$3$     & $7.457$ & $128$      & $16$     & $0.133989$   &  $28$ & $0$ \\
\hline
$1.5$   & $6.872$ & $128$      & $16$     & $0.134986$   & $101$ & $59$ \\
\hline
\hline
\end{tabular}
\end{center}
\caption{
Simulation parameters on lattices.
See the text for details.
}
\label{table:param}
\end{table}

Excitation properties of the quark field are encoded in 
the quark spectral function $\rho_{\mu\nu}(\omega, \bm{p})$, 
with $\mu$ and $\nu$ denoting Dirac indices.
In order to extract $\rho_{\mu\nu}(\omega,\bm{p})$ from lattice 
QCD simulations we have analyzed the quark correlator 
in Euclidean space
\begin{align}
S_{\mu\nu}( \tau,\bm{p} )
= \frac1V \int d^3x d^3y\ 
\e^{ i {\bf p} \cdot ( {\bf x}-{\bf y} ) }
\langle \psi_\mu( \tau,\bm{x} ) \bar\psi_\nu ( 0,\bm{y} ) 
\rangle,
\label{eq:S}
\end{align}
on the lattice in the quenched approximation. 
This correlator is related to the spectral function as
\begin{align}
S_{\mu\nu}( \tau,\bm{p} )
= \int_{-\infty}^\infty d\omega
\frac{ \e^{ (\tau T-1/2) \omega/T }}{ \e^{\omega/2T} + \e^{-\omega/2T} }
\rho_{\mu\nu}( \omega,\bm{p} ).
\label{eq:Stau-rho}
\end{align}
Here, $\tau$ is the imaginary time restricted to the interval 
$0\le\tau<1/T$, $\psi_\mu(\tau,\bm{x})$ is the quark operator, 
and $V$ denotes the volume of the system.
The quark correlator Eq.~(\ref{eq:S}) has been calculated 
after fixing each gauge field configuration to Landau gauge, 
$\partial_\mu A^\mu=0$. 
In the numerical analysis, we used 
$(1/3){\rm tr}|\partial_\mu A^\mu|^2 < 10^{-12}$
as a stopping criterion of the gauge-fixing algorithm.
The simulation parameters are summarized in 
Table~\ref{table:param}.
We used lattices of size $N_\sigma^3\times N_\tau=128^3\times16$
for $T=1.5T_c$ and $3T_c$.
For the lattice fermion, we use non-perturbatively improved 
clover Wilson fermions with the clover coefficient $c_{\rm SW}$
taken from Ref.~\cite{KK09}.
The number of configurations, $N_{\rm conf}$, 
analyzed in the present study for $T=3T_c$ is smaller than 
those in the previous work with $N_{\rm conf}=44-60$.
With the aid of the definition of quark correlator with 
wall source \cite{KK09}, however, the statistics of the 
correlator improves compared to our previous 
analysis, as will be discussed in Sec.~\ref{sec:spectra}.

For $T=1.5T_c$, we find that the quark correlators on 
about half of the gauge configurations show an enhancement 
near the chiral limit deviating from the generic 
behavior \cite{KK09B}. 
As described in detail in the next section,
this enhancement is obviously related to the non-trivial 
topology of gauge field configurations.
The number of configurations which are identified to be
topologically non-trivial is given in Table~\ref{table:param} 
as $N_{\rm top}$.
We found that the influence of topology on the quark propagator
manifest itself in the quark correlator in scalar, pseudo-scalar, 
and tensor channels, whereas vector and axial-vector channels are 
not affected.
Since all analyses in this study depend only on
the quark correlator in the vector channel except for 
the analysis of the quark mass dependence 
given in Fig.~\ref{fig:N_dep}, which shows the 
result only for $T=3T_c$, we have performed 
these analyses using all gauge configurations without
distinction of topologically trivial and non-trivial ones.
For $T=3T_c$ all configurations were identified to be 
topologically trivial.

In the Wilson fermion formulation the quark mass, $m_0$, is 
controlled by the hopping parameter, $\kappa$. 
When we discuss the $m_0$ dependence of the quark spectrum
in what follows, we use the relation
\begin{align}
m_0 = \frac1a \log\left( 1 + \frac12 
\left( \frac1\kappa - \frac1{\kappa_c} \right) \right),
\label{eq:m_0}
\end{align}
which defines the pole mass of the Wilson fermion 
propagator. In the free case, corresponding to 
$\beta\rightarrow \infty$, one has $\kappa_c=1/8$.
At finite values of the gauge coupling, $\beta < \infty$,
the quark mass receives an additive 
renormalization. In this work we determine $\kappa_c$ from 
the behavior of the quark correlation function \cite{KK09},
which is presented in Table~\ref{table:param}.

\section{Quark correlator on topologically-nontrivial configurations}
\label{sec:topology}

In the analysis of the quark correlator we found that
for $T=1.5T_c$ the correlator near the chiral limit shows
a large enhancement deviating from the generic behavior
\cite{KK09} on about half the gauge configurations analyzed
in this study. At $T=3T_c$, on the other hand, no such 
effect has been observed.

In this section, we summarize the behavior of the quark
correlator on these configurations.
We show that characteristic features of this 
enhancement of the quark correlator are consistent
with expectations for the behavior of a quark propagator
on gauge field configurations with non-trivial topology
which give rise to zero modes of the Dirac operator
with definite chirality \cite{'tHooft:1976fv}.

\begin{figure}[tbp]
\begin{center}
\includegraphics[width=.49\textwidth]{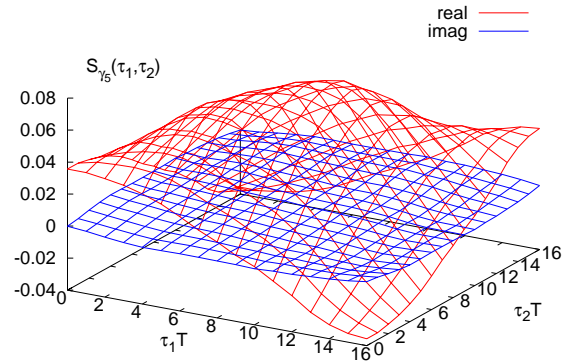}
\caption{
Real and imaginary parts of quark correlator in the 
pseudo-scalar channel, $S_{\gamma_5}(\tau_1,\tau_2)$, on a 
topologically-nontrivial gauge configuration for $T=1.5T_c$.
}
\label{fig:S_P}
\end{center}
\end{figure}

\begin{figure}[tbp]
\begin{center}
\includegraphics[width=.49\textwidth]{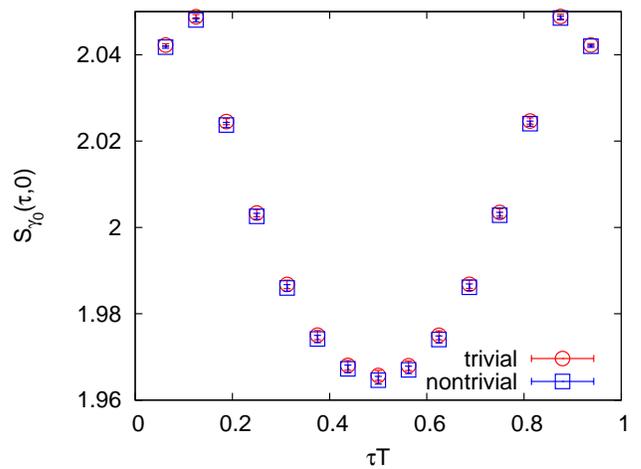}
\caption{
Quark correlators in the temporal component of vector channel, 
$S_{\gamma_0}(\tau)$, averaged over topologically trivial and 
non-trivial gauge configurations for $T=1.5T_c$.
}
\label{fig:S_0}
\end{center}
\end{figure}

The Dirac structure of the quark correlator for $\bm{p}=0$ 
is decomposed into its different quantum number channels
\begin{align}
S_\Gamma(\tau_1,\tau_2)
= \frac1{4V} \int d^3xd^3y {\rm Tr}_{\rm D} 
\left[ \Gamma \langle \psi(\tau_1,\bm{x}) 
\bar\psi(\tau_2,\bm{y}) \rangle \right],
\label{eq:decomp}
\end{align}
where ${\rm Tr}_{\rm D}$ denotes the trace over 
Dirac indices with $\Gamma = 1$, $\gamma_5$, 
$\gamma_\mu$, $\gamma_\mu\gamma_5$, and 
$i[\gamma_\mu,\gamma_\nu]/2$ for scalar, pseudo-scalar, 
vector, axial-vector, and tensor channels, respectively.
In Fig.~\ref{fig:S_P} we show the quark correlator 
in the pseudo-scalar channel, $S_{\gamma_5}(\tau_1,\tau_2)$, 
in the vicinity of the chiral limit on a configuration which 
is identified to be topologically non-trivial.
The quark correlator in the pseudo-scalar channel should vanish
when the system is invariant under parity transformation.
In fact, on all configurations for $T=3T_c$, we have checked
that $S_{\gamma_5}(\tau_1,\tau_2)$ is negligibly small 
compared to scalar and vector channels.
For $T=1.5T_c$, however, on about half of the configurations 
$S_{\gamma_5}(\tau_1,\tau_2)$ shows an enhancement as in 
Fig.~\ref{fig:S_P} near the chiral limit.
The amplitude of this behavior rises with increasing
$\kappa$, i.e., decreasing quark mass, and takes the largest 
value for $\kappa$ near the critical value $\kappa_c$, at which 
the sign is flipped.
A similar behavior is observed in scalar and tensor
channels on these configurations.
On the other hand, such an enhancement does not appear 
in vector and axial-vector channels.
As an example, in Fig.~\ref{fig:S_0} we show the temporal 
component of the quark correlators in the vector channel 
with $\bm{p}=0$, $S_{\gamma_0}(\tau,0)$, in the vicinity of 
the chiral limit averaged separately over topologically trivial 
and non-trivial configurations for $T=1.5T_c$.
The figure shows that the vector-channel correlator on 
each set of configurations does not give a 
statistically-significant difference.

\begin{table}
\begin{center}
\begin{tabular}{ccccc}
\hline
total & $N_0$ & $N_L$ & $N_R$  & $N_{LR}$ \\
\hline
$101$ & $42$  & $14$  & $40$  & $5$ \\
\hline
\end{tabular}
\end{center}
\caption{
Number of configuration for $T=1.5T_c$ 
having no anomalous behavior ($N_0$), 
having in left- and right-handed channels 
($N_L$ and $N_R$, respectively), 
and in both channels ($N_{LR}$). 
}
\label{table:N}
\end{table}

Our numerical result also shows that on many configurations 
which are identified to be topologically non-trivial 
the enhancement manifests itself only on either
the left- or right-handed correlator, which are defined as
\begin{align}
S_{\rm L,R}(\tau,\bm{p}) 
= \frac12 {\rm Tr}_{\rm D} 
[ \Lambda_\pm S(\tau,\bm{p}) ] ,
\label{eq:S_LR}
\end{align}
with projection operators onto left- and right-handed 
quarks, $\Lambda_\pm=(1\pm\gamma_5)/2$.
In Table~\ref{table:N}, we show the number of 
configurations having the enhancement in the left- and 
right-handed channels ($N_L$ and $N_R$), respectively, 
and without the enhancement ($N_0$).
On some configurations an enhancement of the correlator is
observed both in left- and right-handed channels. 
Their number is denoted by $N_{LR}$ and is also given
in Table~\ref{table:N}.

These results on the enhancement of the quark correlator 
near the chiral limit strongly indicate that they are related 
to a zero mode of the Dirac operator which, for instance, may be 
associated with the
presence of instantons on these configurations.
Using the eigenvalues and eigenmodes of 
the Dirac operator $D=\gamma_\mu(\partial_\mu+igA_\mu)$,
\begin{align}
D \psi_\lambda(x) = \lambda \psi_\lambda(x),
\end{align}
the quark propagator, which is the inverse of $iD+m$, 
is written as
\begin{align}
S(x,y) 
= \sum_\lambda 
\frac 1{i\lambda+m} \psi_\lambda(x) 
\psi_\lambda^\dagger(y),
\label{eq:S(x,y)}
\end{align}
where $x$ and $y$ represent positions in Euclidean space.
On a gauge field configuration having an instanton, 
the Dirac operator has a zero mode $\psi_0$ 
\cite{'tHooft:1976fv}, and the quark propagator is 
decomposed as
\begin{align}
S(x,y) 
&= \frac1m \psi_0(x) \psi_0^\dagger(y)
+ \sum_{\lambda \ne 0}
\frac 1{i\lambda+m} \psi_\lambda(x) \psi_\lambda^\dagger(y)
\nonumber \\
&\equiv S_{\rm I}(x,y) + S_\lambda(x,y),
\label{eq:S_lambda}
\end{align}
with 
\begin{align}
S_{\rm I}(x,y) = \frac1m \psi_0(x) \psi_0^\dagger(y).
\label{eq:S_I}
\end{align}
The zero mode associated with an instanton belongs to 
the left-handed spinor and satisfies 
$\Lambda_+\psi_0=\psi_0$.
Accordingly, $S_{\rm I}$ satisfies 
\begin{align}
S_{\rm I}(x,y) = \Lambda_+ S_{\rm I}(x,y) \Lambda_+.
\label{eq:LS0L}
\end{align}
On a configuration having an anti-instanton, 
the zero mode is right-handed, 
$\Lambda_-\psi_0=\psi_0$, and $\Lambda_+$ in 
Eq.~(\ref{eq:LS0L}) are replaced with $\Lambda_-$.
Eq.~(\ref{eq:LS0L}) shows that the zero mode cannot 
affect vector and axial-vector channels, 
since in these channels $\Gamma$ in Eq.~(\ref{eq:decomp}) 
contains one gamma matrix, and hence 
${\rm Tr}_{\rm D} [ \Gamma \Lambda_\pm ] = 0$.

All these properties of $S_{\rm I}$ agree with the 
behavior of the quark correlator observed on the lattice 
for $T=1.5T_c$ discussed above.
Gauge configurations in $N_L$ and $N_R$ would have one or 
more instantons and anti-instantons, respectively, and those 
in $N_{LR}$ have both an instanton and an anti-instanton.
We have also checked that the quark correlator near the 
chiral limit on topologically non-trivial 
configurations approximately satisfies
\begin{align}
S_\Gamma(\tau_1,\tau_2) = \left[ S_\Gamma(\tau_2,\tau_1)\right]^*,
\end{align}
for pseudo-scalar and tensor channels 
(an example is given in Fig.~\ref{fig:S_P}),
which is consistent with Eq.~(\ref{eq:S_I}).

Assuming that the enhancement of the quark correlator indeed
arises from configurations with non-zero topological charge
one can estimate the topological susceptibility 
at that value of the temperature
\begin{align}
\chi_Q = \frac{T\langle \delta Q^2 \rangle }{ V },
\label{eq:chi_Q}
\end{align}
where $Q = n_+ - n_-$ is the topological charge with 
$n_\pm$ being the numbers of instantons and 
anti-instantons, respectively, and $V T^{-1}$ is the 
four-volume in Euclidean space.
Provided that each configuration in $N_L$ and $N_R$ contains 
only one instanton and anti-instanton, respectively, and 
$Q=0$ on configurations in $N_{LR}$, one obtains 
$\langle \delta Q^2 \rangle = ( N_L + N_R )/N_{\rm conf}
\simeq 0.53$ for $T=1.5T_c$.
Using this result in Eq.~(\ref{eq:chi_Q}) with $VT^3= 8^3$ 
we obtain an estimate for
the topological susceptibility of 
$\chi_Q^{1/4} \simeq 0.11 T$ 
for this temperature.
We emphasize that this is a rough estimate since 
the value of $\langle \delta Q^2 \rangle$ alters if some 
topologically non-trivial configurations have more than 
one (anti-)instanton.
Moreover, while the separation between gauge configurations
is a few times larger than the autocorrelation length
measured in terms of the plaquette and Polyakov loop
correlation functions \cite{KK09B}, this may not
be the case for topology and the 
gauge configurations may not be separated 
well enough in Monte Carlo time to suppress a long 
auto-correlation of $Q$.
It is, however, notable that the value of $\chi_Q$ estimated 
here seems to agree with the previous study of $\chi_Q$
on quenched lattice for non-zero $T$ \cite{Gattringer:2002mr}.

The influence of non-trivial topological structures on the 
quark correlator is expected to be non-vanishing 
also in the thermodynamic limit on the quenched lattice.
First, we have numerically checked that the effect does 
not cancel out on a single configuration when averaging
correlation functions over the temporal direction,
\begin{align}
\bar{S}(\tau) = \frac1{N_\tau} \sum_{\tau'} S( \tau+\tau',\tau' ).
\end{align}
Second, the numbers of (anti-)instantons, $n_\pm$, increase 
proportional to $V$, while $\langle|\delta Q|\rangle$ is 
proportional to $V^{1/2}$ \cite{Schafer:1996wv}.
This insures that the topological susceptibility is well
defined in the thermodynamic limit.
Third, instantons and anti-instantons affect different 
channels in the quark correlator, respectively, and the 
numerical result indicates that their effect has a 
definite sign in each channel.
Therefore, effects of topology do not cancel out by 
their average in the thermodynamic limit.
The non-trivial topological properties of
the quark propagator observed here thus will give rise 
to physical contributions in various observables, 
such as those observed as $U(1)_A$ anomaly.

Although the non-trivial topology of gauge configurations gives
rise
to non-vanishing contributions to the quark correlator, 
our numerical result for $T=1.5T_c$ shows that the quark 
propagators in the vector and axial-vector channels are not 
affected by this effect as shown in Fig.~\ref{fig:S_0}.
Since all analyses in Sec.~\ref{sec:spectra} and 
appendix~\ref{sec:S-rho} rely only on the vector channel, 
except for the one for Fig.~\ref{fig:N_dep}, 
we use all gauge configurations including 
topologically trivial and non-trivial ones
in these analyses.
>From the insensitivity of the vector and axial-vector 
channel correlators on topology, it is anticipated that 
the instanton distribution for $T=1.5T_c$ is sufficiently dilute 
so that each instanton affects the quark propagator as 
$S_{\rm I}$ in Eq.~(\ref{eq:S_lambda}) almost independently 
\cite{Schafer:1996wv}.
The analysis on the thermal mass and the dispersion 
relation performed in Sec.~\ref{sec:spectra} would then 
be interpreted as the one for the remaining part $S_\lambda$.

The discussion on the relation between the topology of gauge 
configurations and the enhancement of the quark correlator 
given in this section clearly addresses an important aspect
of QCD thermodynamics and requires further analysis.
We intend to explore the interplay of topology, zero modes, 
eigenfunctions and the structure of the quark propagator in a 
forthcoming publication \cite{prep}.

\section{Quark excitation spectra}
\label{sec:spectra}

\subsection{Quark spectral function and two-pole ansatz}
\label{sec:2pole}

The Dirac structure of the quark spectral function 
$\rho_{\mu\nu}(\omega,\bm{p})$ at finite temperature is 
decomposed as
\begin{align}
\lefteqn{ \rho_{\mu\nu}( \omega, \bm{p} )}
\nonumber \\
&= \rho_0( \omega,p ) (\gamma^0)_{\mu\nu}
- \rho_{\rm v}( \omega,p ) (\hat{\bm{p}}\cdot\bm{\gamma})_{\mu\nu}
+ \rho_{\rm s}( \omega,p ) \bm{1}_{\mu\nu},
\label{eq:rho_0vs}
\end{align}
where $p=|\bm{p}|$ and $\hat{\bm{p}}=\bm{p}/p$ \cite{LeBellac}.
In this section we consider the spectral function above $T_c$
for two cases; (1) for zero momentum, and 
(2) in the chiral limit \cite{KK09}.
With $p=0$, $\rho_{\rm v}( \omega,p )$ vanishes in 
Eq.~(\ref{eq:rho_0vs}) and $\rho_{\mu\nu}( \omega,\bm{p}=\bm{0} )$ 
is decomposed with the projection operators 
$L_\pm = ( 1 \pm \gamma^0 )/2$ as 
\begin{align}
\rho( \omega, \bm{0} )
= \rho^{\rm M}_+( \omega ) L_+ \gamma^0 
+ \rho^{\rm M}_-( \omega ) L_- \gamma^0,
\label{eq:rho^M}
\end{align}
with 
\begin{align}
\rho^{\rm M}_\pm(\omega) 
= \frac12 {\rm Tr}_{\rm D} \left[ \rho(\omega,\bm{0}) \gamma^0 L_\pm \right].
\end{align}
In the chiral limit and for $T>T_c$, the system possesses 
the chiral symmetry and $\rho_{\rm s}( \omega,p )$ vanishes.
$\rho_{\mu\nu}( \omega,\bm{p} )$ is then 
decomposed with the projection operators 
$P_\pm (\bm{p})= ( 1 \pm \gamma^0\hat{\bm{p}}\cdot\bm{\gamma} )/2$
as 
\begin{align}
\rho( \omega,\bm{p} )
= \rho^{\rm P}_+( \omega,p ) P_+(\bm{p}) \gamma^0
+ \rho^{\rm P}_-( \omega,p ) P_-(\bm{p}) \gamma^0,
\label{eq:rho^P}
\end{align}
with 
\begin{align}
\rho^{\rm P}_\pm(\omega,p) 
= \frac12 {\rm Tr}_{\rm D} \left[ 
\rho(\omega,\bm{p}) \gamma^0 P_\pm(\bm{p}) \right].
\end{align}
Using the charge conjugation symmetry one can show that 
\begin{align}
\rho^{\rm M}_\pm(\omega) = \rho^{\rm M}_\mp(-\omega), \quad
\rho^{\rm P}_\pm( \omega,p ) = \rho^{\rm P}_\mp(-\omega,p ).
\label{eq:rhoC}
\end{align}
In the chiral limit and for $p=0$, both $\rho_{\rm s}( \omega,p )$ 
and $\rho_{\rm v}( \omega,p )$ vanish and one obtains
\begin{align}
\rho^{\rm M}_\pm( \omega ) = \rho^{\rm P}_\pm( \omega,0 ) 
= \rho_0(\omega,0).
\label{eq:rho0}
\end{align}

In order to extract the quark spectral function from 
lattice correlator, we follow the approach taken in 
Refs.~\cite{KK07,KK09}, which makes use of a two-pole ansatz 
for the spectrum,
\begin{align}
\rho^{\rm M,P}_+(\omega)
&= Z_1 \delta( \omega - E_1 ) + Z_2 \delta( \omega + E_2 ).
\label{eq:2pole}
\end{align}
Here, $Z_{1,2}$, and $E_{1,2}>0$, are fitting parameters that
will be determined from correlated fits to the lattice 
correlator: $Z_{1,2}$ and $E_{1,2}$ represent the 
residues and positions of poles, respectively. 
By comparing the fit results with spectral functions obtained 
in perturbative calculations one can identify the pole at 
$\omega=E_1$ to be the normal mode, while the one at 
$\omega=-E_2$ corresponds to the plasmino mode 
\cite{BBS92,KKN06,KK09}.
To determine the fit parameters 
with correlated fits, we use lattice data points at
$\tau_{\rm min} \le \tau/a \le N_\tau -\tau_{\rm min}$ 
with the number of points 
$N_{\rm data} = N_\tau-2\tau_{\rm min}+1$.

\begin{figure}[tbp]
\begin{center}
\includegraphics[width=.49\textwidth]{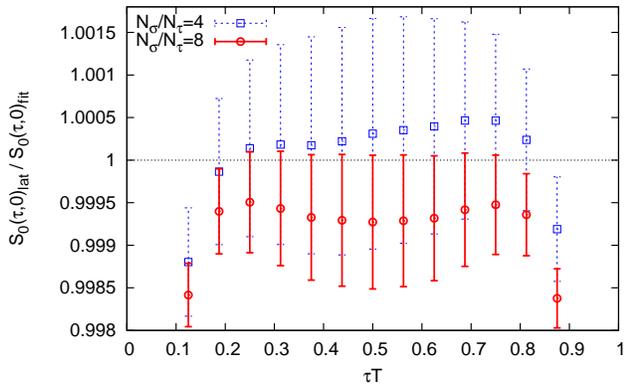}
\caption{
Lattice correlator $S_0(\tau)_{\rm lat}$
in the chiral limit normalized by the fitting function 
Eq.~(\ref{eq:2pole0}) for $T/T_c=3$.
Solid and dashed lines represent the results for 
$N_\sigma/N_\tau=8$ and $4$, respectively.
}
\label{fig:corr}
\end{center}
\end{figure}

\begin{figure}[tbp]
\begin{center}
\includegraphics[width=.49\textwidth]{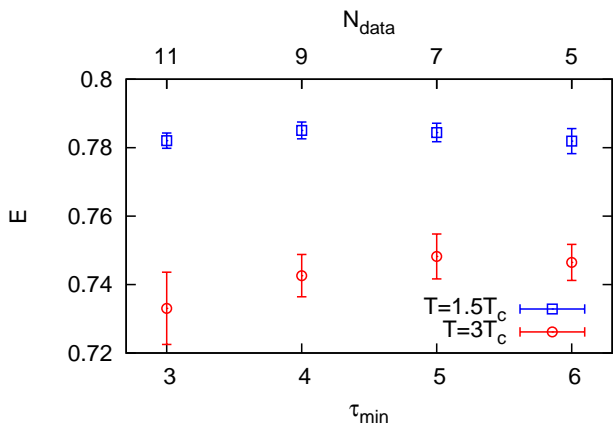}
\caption{
$\tau_{\rm min}$ dependence of fitting parameter $E$ in the
fit with Eq.~(\ref{eq:2pole0}) to
$S_0(\tau,\bm{0})_{\rm lat}$ in the chiral limit.
}
\label{fig:tau_dep}
\end{center}
\end{figure}

Before discussing the analysis of the quark spectrum with this 
fitting function, let us examine the behavior of the quark 
correlator, Eq.~(\ref{eq:S}), on the lattice and discuss how the 
fitting function Eq.~(\ref{eq:2pole}) reproduces it.
To simplify the argument we limit our attention for the 
moment to the case in the chiral limit with $p=0$. 
The Dirac structure of the spectral function is then 
proportional to $\gamma^0$, 
$\rho(\omega,\bm{0}) = \rho_0(\omega,0)\gamma^0$, 
and the Euclidean correlator is solely given by the one 
corresponding to this channel, 
$S_0(\tau,0) = {\rm Tr}_{\rm D} [ S(\tau,\bm{0}) \gamma_0]$.
Since $\rho_0(\omega,0)$ is even in $\omega$, 
the ansatz Eq.~(\ref{eq:2pole}) reduces to 
\begin{align}
\rho_0 ( \omega,0 )
&= \rho^{\rm M}_+(\omega)
= \rho^{\rm P}_+(\omega,0) \nonumber \\
&= Z \left( \delta( \omega - E ) + \delta( \omega + E ) \right),
\label{eq:2pole0}
\end{align}
which involves only two fit parameters, $Z$ and $E$.
In Fig.~\ref{fig:corr} we show the lattice correlator
$S_0(\tau,0)_{\rm lat}$ for $T/T_c=3$ normalized 
by the one constructed from the fitting function 
Eq.~(\ref{eq:2pole0}), $S_0(\tau,0)_{\rm fit}$, with fitting 
parameters determined from a correlated fit with $\tau_{\rm min}=4$.
The figure shows that the fit result, $S_0(\tau,0)_{\rm fit}$, well 
reproduces the lattice results, $S_0(\tau,0)_{\rm lat}$, for 
$\tau T \simeq 0.5$.
We have checked that the fitting parameters do not have a
statistically significant dependence on the choice of 
$\tau_{\rm min}$ for $3\le\tau_{\rm min}\le6$. 
This is evident from the dependence of $E$
on $\tau_{\rm min}$.
Results obtained in the chiral limit 
are shown in Fig.~\ref{fig:tau_dep}. 
In Fig.~\ref{fig:corr},
clear deviations between $S_0(\tau,0)_{\rm fit}$ and 
$S_0(\tau,0)_{\rm lat}$ are observed 
for small and large $\tau$, 
which can be attributed to distortion effects 
arising from the source \cite{KK09,distortion}.
For $\tau_{\rm min}\le2$, $\chi^2/{\rm dof}$ of the fit 
becomes unacceptably large due to the 
distortion effect.
Similar results are obtained for all parameters 
analyzed in this study with Eq.~(\ref{eq:2pole}).
In the following, we use $\tau_{\rm min}=4$, which is chosen in 
order to reduce the distortion effect and at the same time 
leave us with a sufficient number of data points for our four 
parameter fits.
We found that with $\tau_{\rm min}=4$ fits with the pole 
ansatz give reasonable values of chi-square, 
$0.5<\chi^2/{\rm dof}<3$, for $m_0/T\lesssim 0.5$ and
$p/T\lesssim 3$.

We remark that, although the two-pole ansatz 
Eq.~(\ref{eq:2pole}) reproduces the lattice correlator quite well
with a small $\chi^2/{\rm dof}$, this result does not necessarily 
mean that the quark spectral function is solely composed 
of two sharp peaks at energies $\omega=E_1$ and $-E_2$.
In fact, as discussed in appendix~\ref{sec:S-rho}, the 
Euclidean correlator is insensitive to the structure of the 
spectral function 
at $|\omega|\lesssim T$. 
Although with our current statistics the quark propagator 
at all distances is determined to better than $0.1\%$, the  
spectrum in this low energy range generally still suffers
from large uncertainties.
In Sec.~\ref{sec:p>0} and appendix~\ref{sec:S-rho}, we will 
discuss how our analysis of the quark spectrum 
is influenced by this problem.
Nevertheless, we will assert in these sections that, from 
the success of the two-pole ansatz over wide ranges of 
$m_0$ and $p$, it is reasonable to expect that the 
quark spectrum contains peak structures corresponding to 
the normal and plasmino modes.
Qualitative dependences of the positions of these peaks 
on $m_0$ and $p$ are then analyzed with 
the fitting parameters $E_1$ and $E_2$.

Finally, we briefly comment on the statistical errors
of the quark correlator in this analysis.
In Fig.~\ref{fig:corr}, the quark correlator 
obtained in Ref.~\cite{KK09} with $N_\sigma/N_\tau=4$ 
for $T=3T_c$ is shown by the dotted lines together with the 
present result with $N_\sigma/N_\tau=8$.
The figure shows that the error-bars for 
$N_\sigma/N_\tau=8$ is about $40\%$ smaller in magnitude 
than that for $N_\sigma/N_\tau=4$, although the former 
result is obtained with only about half as many 
configurations than the latter ($N_{\rm conf}=51$). 
As mentioned previously, this improvement in statistics 
is a consequence of the use of the wall source in the 
numerical analysis of quark correlators \cite{KK09}.
We have checked that the distribution of correlators
obtained on each gauge configuration around the 
average is approximately proportional to $V^{-1/2}$. 
Such a scaling with the wall source is to be expected 
when the spatial extent of the lattice, $L$, is larger 
than the coherence length of the quark field, $\xi$.
Estimating $\xi T \simeq T/m_T \simeq 1.25$ and 
$LT = N_\sigma/N_\tau =8$ this, indeed, seems to hold.

\subsection{Excitation spectra for \boldmath $p=0$}
\label{sec:m>0}

\begin{figure}[tbp]
\begin{center}
\includegraphics[width=.49\textwidth]{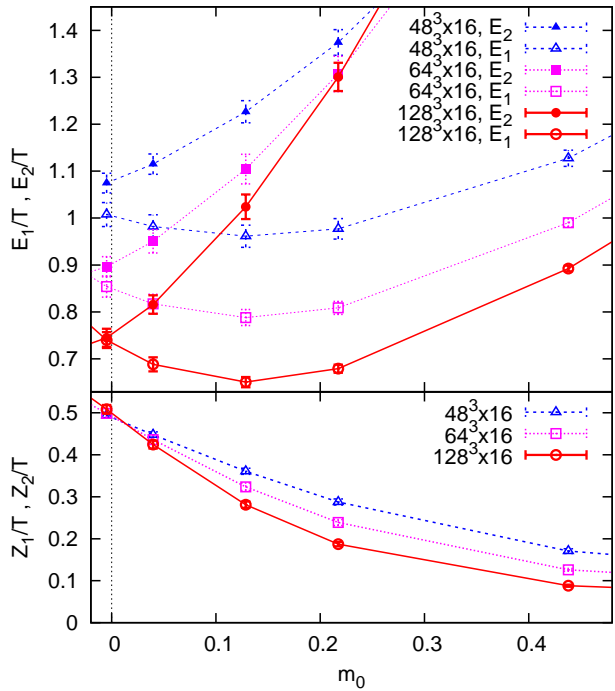}
\caption{
Bare quark mass dependence of parameters $E_1$, $E_2$ at
$T=3T_c$ on lattices with $N_\sigma/N_\tau=8$, $4$ and $3$.
}
\label{fig:N_dep}
\end{center}
\end{figure}

\begin{table}
\begin{center}
\begin{tabular}{c|cccc}
\hline
\hline
$N_\sigma/N_\tau$ & $3$      & $4$    & $8$   & $\infty$ \\
\hline
$T/T_c=3$    & $1.041(16)$ & $0.875(8)$ & $0.743(6)$ & $0.725(14)$  \\
$T/T_c=1.5$  & $1.075(14)$ & $0.906(8)$ & $0.784(3)$ & $0.768(11)$  \\
\hline
\hline
\end{tabular}
\end{center}
\caption{
The values of $m_T/T$ obtained on lattices with 
$N_\sigma/N_\tau=3$, $4$ and $8$.
The far right column shows the values of $m_T$ estimated 
by the extrapolation to the infinite volume limit.
}
\label{table:m_T}
\end{table}

In this subsection, we focus on the quark spectrum with 
$p=0$, which is decomposed into $\rho^{\rm M}_\pm(\omega)$ 
as in Eq.~(\ref{eq:rho^M}), and analyze the $m_0$ dependence 
of the spectrum.
In Fig.~\ref{fig:N_dep}, we show results for the $m_0$ 
dependence of fitting parameters, 
$E_1$, $E_2$, and $Z_2/(Z_1+Z_2)$ for $T=3T_c$. 
We compare the results on lattices with smaller aspect 
ratios $N_\sigma/N_\tau=3$ (triangles) and $4$ (squares) 
\cite{KK07,KK09} with our new results 
on the $128^3\times16$ lattice (circles).
One sees that the qualitative behavior of fitting parameters
on the largest lattice agrees with previous 
results obtained on lattices with smaller $N_\sigma/N_\tau$.
The ratio $Z_2/(Z_1+Z_2)$ becomes larger as $m_0$ is 
decreased, and eventually reaches $Z_2/(Z_1+Z_2)=0.5$
in the chiral limit.
At this point, which defines $\kappa_c$ in Eq.~(\ref{eq:m_0}), 
$E_1=E_2$ is satisfied within the statistical error.
The thermal mass is defined as $m_T \equiv ( E_1 + E_2 )/2 $ 
on this point, or equivalently $m_T \equiv E$ with the fit 
with Eq.~(\ref{eq:2pole0}) to $S_0(\tau,0)$.
For $m_0>0$, $E_1$ has a minimum, while 
$E_2$ is a monotonically increasing function.
For $T=1.5T_c$, the same analysis can be performed on 
topologically trivial gauge configurations, which 
gives qualitatively the same result 
as in Fig.~\ref{fig:N_dep}.

Figure~\ref{fig:N_dep} also shows that the values of $E_1$ 
and $E_2$ drop significantly with increasing aspect ratio
$N_\sigma/N_\tau$. 
Accordingly, the value of $m_T$ also has a strong 
$N_\sigma/N_\tau$ dependence.
The value of $m_T/T$ for $N_\sigma/N_\tau=3$, $4$ and $8$ 
for $T/T_c=1.5$ and $3$ is given in Table~\ref{table:m_T}.
To infer the thermal mass in the infinite volume limit, 
we performed an extrapolation of $m_T$ to infinite volume 
with an ansatz 
\begin{align}
m_T(1/V) \sim m_T(0) \exp(c/V)
\end{align}
using the results with $N_\sigma/N_\tau=4$ and $8$.
The result of this extrapolation is shown in 
the far right column of Table~\ref{table:m_T}.
The infinite volume extrapolated values of $m_T$ for each $T$ 
obtained in this study
is about $10\%$ smaller in magnitude than those estimated 
in Ref.~\cite{KK09}.
It is also notable that the ratio $m_T/T$ increases 
as $T$ is lowered, which was not observed in the previous 
study \cite{KK09}.
With the present extrapolation, the value of $m_T$ in the 
infinite volume limit coincides with the one obtained 
for $N_\sigma/N_\tau=8$ 
within the statistical error. This result suggests that 
the influence of finite volume effects on $m_T$ is well 
suppressed for $N_\sigma/N_\tau=8$.

In Fig.~\ref{fig:N_dep}, one also finds that the value of 
$\kappa_c$, {\it i.e.} $m_0$ at which $Z_2/(Z_1+Z_2)=0.5$, 
slightly shifts with the variation of $N_\sigma/N_\tau$.
The difference of $\kappa_c$ between the two largest 
lattices $N_\sigma/N_\tau=4$ and $8$, however, is within 
the statistical error.

\subsection{Dispersion relations at non-zero $p$}
\label{sec:p>0}

\begin{figure}[tbp]
\begin{center}
\includegraphics[width=.49\textwidth]{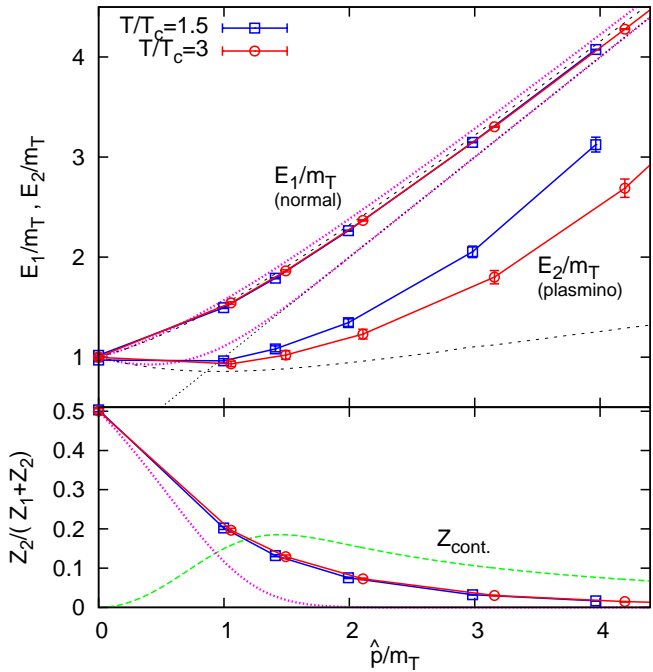}
\caption{
Dependences of the fitting parameters 
$E_1$ and $E_2$ and the ratio $Z_2/(Z_1+Z_2)$ on
the lattice momentum $\hat{p}=(1/a) \sin(pa)$
for $T/T_c=1.5$ and $3$.
See, the text for the explanation of other lines.
}
\label{fig:disp}
\end{center}
\end{figure}

Next, we set $\kappa=\kappa_c$ and analyze the momentum 
dependence of the excitation spectra at non-zero momentum 
in the chiral limit using the decomposition in 
Eq.~(\ref{eq:rho^P}).
For the analysis of the quark correlator at non-zero $p$ 
for $T=1.5T_c$, we have performed the analysis on $54$ 
configurations including topologically trivial and 
non-trivial ones.
In Fig.~\ref{fig:disp} we show the momentum dependence 
of $E_1$ and $E_2$ normalized by $m_T$, as well as 
$Z_2/(Z_1+Z_2)$, for $T/T_c=1.5$ and $3$.
The horizontal axis represents the momentum of a free 
fermion on the lattice, $\hat{p}=(1/a) \sin pa$, 
normalized by $m_T$.
The figure shows that the qualitative features of the
dispersion relations as functions of $\hat{p}$ do not change
compared to our earlier results on lattices with smaller 
aspect ratio \cite{KK09}:
$E_1>E_2$ is always satisfied in contrast with the results 
in the previous subsection, and $E_2$ enters the space-like 
region at high momentum.
We note that a similar result is obtained from the analysis 
of the quark spectrum in the Schwinger-Dyson approach 
\cite{Mueller:2010ah,Qin:2010pc}.
The decrease of $Z_2/(Z_1+Z_2)$ for large momentum shows 
that the plasmino mode ceases to exist as $p$ becomes larger.

At asymptotically high temperatures the quark spectrum 
is calculated perturbatively \cite{HTL,LeBellac}.
The dispersion relations of the normal and plasmino modes 
obtained at the one-loop order are shown in Fig.~\ref{fig:disp} 
by the dotted lines with $m_T=gT/\sqrt{6}$.
The residue of the plasmino mode in the normalization of the 
quark field $\int_{-\infty}^{\infty} d\omega 
\rho^{\rm P}_+(\omega,p)=1$ is also shown in the lower panel
by the dotted line. 
Besides these poles, $\rho^{\rm P}_+(\omega,p)$ at the 
one-loop order has a continuum in the space-like region.
The spectral weight of the continuum, 
$Z_{\rm cont.} \equiv \int_{-p}^{p} d\omega 
\rho^{\rm P}_+(\omega,p)$, is shown in the lower panel 
by the dashed line,
which in the perturbative analysis reaches at most
exceeds $Z_{\rm cont.}\simeq0.18$.

The numerical result for $E_2$ presented in Fig.~\ref{fig:disp} 
shows that the result for the lowest non-zero momentum,
$p_{\rm min} = 2\pi (N_\tau/N_\sigma)T$, is significantly lower 
than $m_T$; {\it i.~e.} $m_T/T \simeq E_2/T|_{p=0}=0.742(12)$ 
and $E_2/T|_{p=p_{\rm min}}=0.694(21)$ for $T/T_c=3$ and 
$E_2/T|_{p=0}=0.782(9)$ and $E_2/T|_{p=p_{\rm min}}=0.758(18)$
for $T/T_c=1.5$. 
Provided that the value of $E_2$ in our two-pole ansatz 
represents the dispersion relation of the plasmino mode,
this result serves as direct evidence for the existence of 
the plasmino minimum in the non-perturbative analysis.
One, however, has to be careful with this interpretation.
As we have argued previously and in appendix~\ref{sec:S-rho}, 
the Euclidean correlator is insensitive to the spectral function 
at low energy, $|\omega|\lesssim T$, and analysis of the 
spectrum in this energy range has large uncertainty.
In fact, in terms of the comparison with the perturbative 
result our fitting function does not contain the continuum 
in the space-like region. 

In order to see the effect of the continuum part in the spectral 
function at the one-loop order on the Euclidean correlator,
we have performed the following analysis.
First, we calculated the Euclidean correlator at one-loop 
order in perturbation theory, $S_{\rm HTL}(\tau,\bm{p})$.
We then performed the two-pole fit with Eq.~(\ref{eq:2pole}) 
to discrete points of $S_{\rm HTL}(\tau,\bm{p})$ with the 
same discretization as on the lattice.
To estimate the chi-square of the fit, we put error-bars and 
Gaussian noise on each value of $S_{\rm HTL}(\tau,\bm{p})$ 
following the covariance matrix obtained on the lattice.
With this analysis, we found that the two-pole fit to 
$S_{\rm HTL}(\tau,\bm{p})$ gives a reasonable chi-square, 
$\chi^2/{\rm dof}\simeq1$, although the spectral function
corresponding to $S_{\rm HTL}(\tau,\bm{p})$ contains 
the continuum. 
The values of $E_1$ and $E_2$ obtained by the fit 
to $S_{\rm HTL}(\tau,\bm{p})$ are shown by the 
dashed lines in the upper panel of Fig.~\ref{fig:disp}.
One sees that these results considerably deviate from the 
positions of normal and plasmino modes. In particular, 
we note that $E_2$ also
enters the space-like region for $p \gtrsim T$.
This result obviously shows that the success of the pole 
ansatz does not mean the absence of the continuum.
The values of $E_1$ and $E_2$ thus should not be regarded as 
the positions of peaks, but a center of spectral 
weights in some sense in the positive and negative energy 
regions.
It also suggests that in our current analysis the 
normal mode is quantitatively much better under control
than the plasmino branch.

Nonetheless, some qualitative features on the quark spectrum 
can be extracted from our results. First, as discussed in 
Appendix~\ref{sec:S-rho}, the consistent success of the 
two-pole ansatz over wide parameter ranges suggests that the 
quark spectrum has peak structures at positive and negative 
energies in the range of $m_0$ and $p$ analyzed in this study.
Next, the momentum dependences of the peak positions in 
$\rho^{\rm P}_+(\omega,p)$ can be constrained by general 
properties of the spectral function.
Using the relation Eq.~(\ref{eq:rhoC}) and the analyticity 
of $\rho(\omega,\bm{p})$, 
one can show that (1) if $\rho^{\rm P}_+(\omega,p)$ at $p=0$ 
has a peak at positive energy $\omega=E^*_+(0)$, there also 
exists a peak at negative energy with the same absolute
value, $\omega=-E^*_-(0)=-E^*_+(0)$, 
and (2) momentum dependences of the peak positions 
$\omega=E^*_+(p)$ and $-E^*_-(p)$ satisfy 
$dE^*_+/dp = -dE^*_-/dp $ at $p=0$.\footnote{
A discussion of a similar relation for position of {\it poles}
in the quark spectral function is presented in \cite{Peshier:1999dt}
The extension of this argument to the position of {\it peaks}
is straightforward. This feature is graphically seen,
for example, in the contour maps of spectral functions
in Ref.~\cite{KKN06}.
}.
The result in Fig.~\ref{fig:disp} indicates not only a
negative $dE_2/dp$ at $p=0$ but also a positive $dE_1/dp$.
It thus seems reasonable to conclude that the dispersion 
relation of the plasmino peak, $E^*_-(p)$, has a 
negative derivative near $p=0$. 
This result then leads to the existence of a minimum of 
$E^*_-(p)$, provided that the plasmino peak survives at 
sufficiently high momentum where the $dE^*_-/dp$ becomes 
positive.

\section{Brief Summary}
\label{sec:discussion}

In this study, we extended the analysis of the quark 
spectral function on quenched lattices, performed in previous work 
\cite{KK07,KK09}, to lattices with larger spatial volume.
With this analysis we achieved considerable progress in the 
understanding of finite spatial volume effects 
on the quark correlator 
and confirmed the importance of non-trivial topological
configurations for the structure of the quark propagator
at temperatures above $T_c$.
The strong $V$ dependence observed in a previous 
study with aspect ratios $N_\sigma/N_\tau=3$ and $4$ seems to 
converge on the largest lattice with $N_\sigma/N_\tau=8$. 
An extrapolation of the quark thermal mass, $m_T$, to the 
infinite volume is about $10\%$ smaller than the earlier 
estimate \cite{KK09}. 

Dependences of the fitting parameters in the two-pole 
ansatz on $m_0$ and $p$ do not change qualitatively 
on the largest lattice.
We, however, found that the position of the pole 
corresponding to the plasmino, $E_2$, has a minimum at 
non-zero momentum within the ansatz.
As discussed in Sec.~\ref{sec:p>0}, this result with the 
pole ansatz does not directly mean the existence of the 
minimum of the plasmino dispersion (See also, 
Ref.~\cite{Mueller:2010ah}.).
The quark correlator obtained on the lattice as functions of 
continuous parameters $m_0$ and $p$, however,
strongly indicates the existence of the plasmino peaks and 
that the dispersion relation of the plasmino has a negative 
slope at low momentum near $T_c$.

We also presented evidence for the influence of non-trivial
topology of gauge field configurations on the structure
of the quark propagator. This clearly is visible in our
data at $T=1.5T_c$, while at $T=3 T_c$ we do not observe
topologically non-trivial configurations. We showed that,
nonetheless, the vector and axial vector channels are not 
influenced by this effect at $T=1.5T_c$.
We expect that the effect 
of non-trivial topology will modify the structure of the
quark propagator in the quenched QCD significantly
as $T$ approaches $T_c$ from above.
The analysis of correlators near $T_c$ therefore 
requires a careful treatment of this effect.

\appendix
\section{Difficulty in analyzing the low energy structure of 
spectral functions}
\label{sec:S-rho}

In this study, we have used the two-pole ansatz Eq.~(\ref{eq:2pole}) 
to examine the quark spectrum from Euclidean lattice correlator.
This ansatz is found to reproduce the lattice correlator 
well over wide parameter ranges. 
In this appendix, we address the significance of the success 
of the two-pole ansatz.
We show that this success to some extent is due to a poor 
resolution of the lattice correlator on the spectral function 
at low energy, $|\omega|\lesssim T$.
The insensitivity of the Euclidean correlator to low-energy
spectrum has been recognized through the studies of 
spectral functions on the lattice, especially efforts to 
extract transport properties 
\cite{Aarts:2002cc,Petreczky:2005nh,Meyer:2009jp}.
Since the spectral weight of quarks with a small mass and 
a momentum concentrates in the energy range $|\omega|\lesssim T$,
our analysis suffers from the same difficulty.

In the next subsection we first summarize the problem of the 
insensitivity in terms of the power series expansion of correlators.
We then take a closer look at the quark correlator
in Sec.~\ref{sec:corr-app}.

\subsection{Moment expansion}
\label{sec:mom-exp}

Let us consider a Euclidean correlator $S(\tau)$ and a 
spectral function $\rho(\omega)$ which are related to 
each other as 
\begin{eqnarray}
S(\tau) = \int_{-\infty}^{\infty} d\omega 
\frac{ e^{ (\tau T - 1/2)\omega/T} }
{ e^{\omega/2T} +\zeta e^{-\omega/2T} } \rho(\omega), 
\label{eq:Srho}
\end{eqnarray}
where $\zeta=\pm1$ for fermions and bosons, respectively.
By Taylor expanding $\exp[(\tau T -1/2)\omega/T]$,
Eq.~(\ref{eq:Srho}) is written as
\begin{align}
S(\tau) = \sum_n C_n \left( \tau T - \frac12 \right)^n,
\label{eq:Taylor}
\end{align}
where the coefficients $C_n$ are given by 
the moments of thermal spectral function, 
$\rho'(\omega) \equiv 
\rho(\omega) / (e^{\omega/2T} +\zeta e^{-\omega/2T} )$,
\begin{align}
C_n = \frac1{n!} \int_{-\infty}^\infty d\omega 
\left(\frac\omega T\right)^n 
\rho'( \omega,p ) .
\label{eq:C_n}
\end{align}
One can show that the power series Eq.~(\ref{eq:Taylor}) 
converges for $0<\tau<1/T$ for 
spectral functions having an asymptotic form 
$\rho(\omega) \sim \zeta\omega^m$ for large $|\omega|$.

In lattice simulations one obtains the values of $S(\tau)$
for $N_\tau$ discrete $\tau$ values with statistical errors.
Through Eq.~(\ref{eq:Srho}), these values of $S(\tau)$ provide 
$N_\tau$ 'pieces of different' information on the structure of 
$\rho(\omega)$.
The coefficients of the power series, $C_n$, can be 
interpreted as the different representation of this 
information.

To consider how the information of low-energy structure 
of $\rho(\omega)$ is encoded in $S(\tau)$,
we decompose $\rho(\omega)$ into low and high energy parts 
as $\rho(\omega) 
= \rho_\Lambda(\omega) + \rho_{\rm high}(\omega)$ with 
$\rho_{\Lambda}(\omega) = \rho(\omega)\theta(\Lambda-|\omega|)$
with some energy $\Lambda$.
Since $S(\tau)$ and $\rho(\omega)$ are related with each 
other linearly, one can then also decompose $S(\tau)$ 
into contributions of low and high energy spectrum.
The contribution of $\rho_\Lambda(\omega)$ is written as
\begin{align}
S_\Lambda(\tau)
\equiv \int d\omega \frac{ e^{ (\tau T - 1/2 )\omega/T} }
{ e^{\omega/2T} +\zeta e^{-\omega/2T} } \rho_\Lambda(\omega)
= \sum_n S^{(\Lambda)}_n(\tau) ,
\end{align}
with
\begin{align}
S^{(\Lambda)}_n(\tau) 
&= C^{(\Lambda)}_n \left(\tau T-\frac12\right)^n,
\label{eq:S^n_low}
\\
C^{(\Lambda)}_n 
&= \frac1{n!}
\int_{-\Lambda}^\Lambda d\omega 
\left(\frac\omega T\right)^n \rho'(\omega).
\label{eq:C^low}
\end{align}
To avoid unnecessary complexity in the following we limit our 
attention to the spectral function satisfying 
$\rho(\omega) = \zeta\rho(-\omega)\ge0$ for $\omega>0$.
Odd terms in Eq.~(\ref{eq:C^low}) then vanish while
$C^{(\Lambda)}_n\ge0$ for even $n$.
With the aid of inequalities $(\omega/T)^n\le (\Lambda/T)^n$ 
in the integrand of Eq.~(\ref{eq:C^low}) and 
$|\tau T-1/2|<1/2$, one then obtains an inequality
\begin{align}
\frac{ S^{(\Lambda)}_n(\tau) }{ S(\tau) }
\le \frac{ S^{(\Lambda)}_n(\tau) }{ S^{(\Lambda)}_0(\tau) }
\le \frac1{n!} \left( \frac\Lambda{2T}\right)^n.
\label{eq:<}
\end{align}
This inequality provides an upper limit of the contribution 
of $S^{(\Lambda)}_n(\tau)$ to $S(\tau)$, {\it i.e.} 
the relative strength of signals for information on each 
moment of $\rho'_\Lambda(\omega)$ encoded in $S(\tau)$.
Eq.~(\ref{eq:<}) tells us that if we choose $\Lambda \lesssim T$ 
the contribution of higher order moments is strongly 
suppressed due to the factorial and exponential factors. 
The decrease of 
the signals makes evaluations of their values difficult. 
In practical analyses with statistical errors, therefore, 
the number of moments, {\it i.e.} the number of independent 
information on $\rho_\Lambda(\omega)$, which can be 
estimated from the correlator is severely limited with 
Eq.~(\ref{eq:<}).
For example, if one wants to examine $\rho_\Lambda(\omega)$
with $\Lambda=T$ from the lattice correlator having 
statistical errors of order $0.01\%$, 
$S^{(\Lambda)}_6(\tau)/S(\tau) \le (1/6!)2^{-6}\simeq 2\times10^{-5}$ 
is already significantly smaller than the statistical error. 
Moments of order higher than the sixth
are inaccessible 
unless the correlators with different $\tau$ values
have strong correlations corresponding to the moments.
One can obtain at most three statistically meaningful
information on $\rho_\Lambda(\omega)$, {\it i.e.} 
$C^{(\Lambda)}_0$, $C^{(\Lambda)}_2$ and $C^{(\Lambda)}_4$, 
from the correlator.
We emphasize that this limitation on the number of independent
information on $\rho_\Lambda(\omega)$ is not resolved 
with the increase of $N_\tau$.

When $\rho_{\rm high}(\omega)$ gives rise to a large 
contribution to $S(\tau)$, the analysis of 
$\rho_\Lambda(\omega)$ should be much more difficult. 
This occurs particularly when $\rho(\omega)$ has a positive 
mass dimension, $m$, since $\rho(\omega)$ behaves as 
$\sim \omega^m$ for large $\omega$ for this case.
Since the quark spectral function has a negative mass
dimension, effects of $\rho_{\rm high}(\omega)$ is
expected to be small in our analysis.

While in the above argument we have limited our attention to 
the spectrum satisfying $\rho(\omega)=\zeta\rho(-\omega)$,
one can also derive a similar inequality for odd terms of 
$S^{(\Lambda)}_n(\tau)$ for general cases.

\subsection{Quark correlator}
\label{sec:corr-app}

\begin{figure}[tbp]
\begin{center}
\includegraphics[width=.45\textwidth]{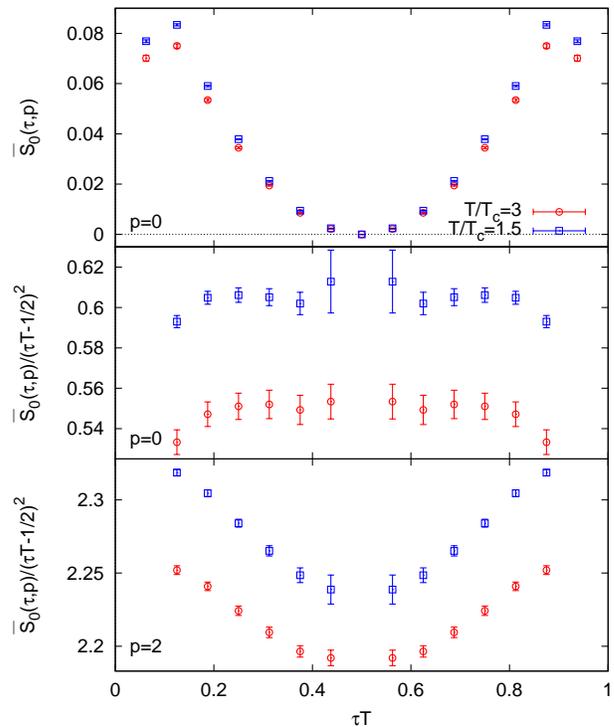}
\caption{
Mid-point subtracted correlator 
$\bar{S}_0(\tau,0)$ in the chiral limit 
(upper panel), and 
$\bar{S}_0(\tau,p) / (\tau T-1/2)^2$
for $p=0$ (middle) and $2(2\pi N_\tau/N_\sigma)T$ (lower).
}
\label{fig:2nd}
\end{center}
\end{figure}

Next, we inspect the structure of the quark correlator 
based on the discussion in the previous subsection.
Let us first consider the correlator in the chiral limit 
with $p=0$.
In this case, the quark correlator takes non-zero values
only in the temporal component of the vector channel, 
$S_0(\tau,0)_{\rm lat}$.
The correlator in this channel is symmetric,
$S_0(\tau,0) = S_0(1/T-\tau,0)$,
and odd terms vanish in Eq.~(\ref{eq:Taylor}).
The values of $C_n$ for even $n$ are evaluated from the 
correlator as follows.
First, the lowest moment, $C_0$, is directly read off from 
the correlator at $\tau=1/(2T)$; 
$ C_0 = S_0( 1/(2T),0 )_{\rm lat} $.
To examine higher order moments, we show
in Fig.~\ref{fig:2nd} the $\tau$ dependence of 
the mid-point subtracted correlator
\begin{align}
\bar{S}_0( \tau,0 ) \equiv S_0(\tau,0) - S_0(1/2T,0),
\end{align}
in the upper panel and that divided by $(\tau T-1/2)^2$ 
in the middle panel.
Using the latter plot, $C_2$ is estimated as 
$C_2 = \lim_{\tau T \to 1/2} \bar{S}_0(\tau,0)/(\tau T-1/2)^2$,
while the deviation from the constant in this plot carries 
information of $C_n$ for $n\ge4$. The figure, however, shows 
that $\bar{S}_0( \tau,0 )/(\tau T-1/2)^2$ is constant within 
the statistical error except for the range of $\tau$ near 
the source, $|\tau T-1/2|\gtrsim0.3$, which receives the 
distortion effect and should be excluded from the analysis.
This result shows that information on moments higher than 
four are buried in the statistical errors and inaccessible 
from our numerical results for this parameter.

This result is consistent with the argument in the 
previous subsection. 
As the pole ansatz predicts the poles of $\rho_0(\omega,0)$ 
at $|\omega|/T\simeq0.74$, spectral weight of 
$\rho_0(\omega,0)$ is expected to concentrate more or less 
in the energy range $|\omega|<T$.
Substituting $\Lambda=T$ to Eq.~(\ref{eq:<}), one obtains 
that $S_n^{(\Lambda)}(\tau,0)_{\rm lat} 
/ S(\tau,0)_{\rm lat}$ 
are suppressed faster than $(2^n n!)^{-1}$.
This upper limit is already the same order as 
the typical magnitude of the statistical error of 
the correlator, $10^{-3}$, for $n=4$.
One thus can obtain only two statistically meaningful 
information on the quark spectrum, $C_0$ and $C_2$, from
the correlator for this parameter.

Since the quark correlator for $m_0=p=0$ contains only 
two information on the quark spectrum, any ansatz having 
more than two fitting parameters can well reproduce the 
correlator as long as it gives sufficiently small higher 
order moments.
For example, we found that a fitting function 
with a single Gaussian peak for $\rho_0(\omega,0)$ 
in the chiral limit,
\begin{align}
\rho_0(\omega,0) = \rho^{\rm M}_+(\omega)
= \frac Z{\sqrt{\pi}\Gamma} \exp 
\left( -\frac{\omega^2}{\Gamma^2} \right),
\label{eq:2gauss0}
\end{align}
including two fitting parameters, $Z$ and $\Gamma$, gives 
$\chi^2/{\rm dof}\simeq2.5$ for $T/T_c=3$, which is less
than twice larger than the one with the two-pole ansatz.

Evaluation of the quark spectrum for $m_0=p=0$ with the 
lattice correlator, therefore, is quite difficult.
It, however, is notable that one can obtain some additional
information on the spectrum for $m_0=p=0$ from the numerical 
analysis by using correlator for non-zero $m_0$ and $p$, 
because the spectral function is continuous with respect to 
these parameters.
When $m_0$ ($p$) takes non-zero values, 
$\rho^{\rm M}_+(\omega)$ ($\rho^{\rm P}_+(\omega,p)$) 
is no longer an even function. Odd moments in 
Eq.~(\ref{eq:Taylor}) then take non-zero values and 
these values provide additional constraints for the spectrum.
At the same time the strength of $\rho^{\rm M,P}_+$ shifts 
toward higher energies, as is expected from the free quark 
spectrum, $E=\sqrt{p^2+m_0^2}$.
The higher order moments can then become larger, which makes 
the estimates of their values easier.
To see this explicitly, in the lower panel of Fig.~\ref{fig:2nd} 
we show $\bar{S}_0(\tau,p)_{\rm lat}/(\tau T-1/2)^2$ 
with $p=2p_{\rm min}=2(2\pi N_\tau/N_\sigma)T$.
The figure shows that this function clearly deviates
from a constant, and hence contains statistically significant 
information on the fourth moment and higher.
The number of information accessible with the quark correlator 
with this parameter thus is at least more than five.
Since this number is larger than that of fitting parameters in 
the two-pole ansatz, the small $\chi^2/{\rm dof}$ with the 
ansatz for this correlator is non-trivial.
Since the simple two-poles ansatz Eq.~(\ref{eq:2pole}) 
can reproduce the lattice correlator not only for $m_0=p=0$ 
but also over wide ranges of continuous parameters, 
it is expected that this ansatz reproduces at least a 
qualitative feature of the global structure of the quark 
spectrum in this parameter range, including $m_0=p=0$.

Similarly, one can argue that the single Gaussian ansatz, 
Eq.~(\ref{eq:2gauss0}), in the chiral limit and for $p=0$
is most probably excluded as follows.
The continuity of the quark spectrum means that if the Gaussian 
ansatz appropriately reproduces the structure of the spectrum, 
the ansatz has to be continuously connected to non-zero $m_0$.
We, however, have checked that several ans\"atze for 
$\rho^{\rm M}_+(\omega)$ which reduce to Eq.~(\ref{eq:2gauss0}) 
at $m_0=0$ cannot give a reasonable chi-square with $m_0\ne0$; 
as far as we have checked, $\chi^2/{\rm dof}$ obtained on
all the ans\"atze grows rapidly as $m_0$ is increased.
Although this analysis cannot completely exclude the 
form of Eq.~(\ref{eq:2gauss0}) in the chiral limit,
it strongly suggests that the quark spectrum like 
Eq.~(\ref{eq:2gauss0}) is most probably ruled out 
even in the chiral limit.

\section*{Acknowledgments}
\label{sec:Ackn}

This work has been supported in part by a Grant-in-Aid for Scientific 
Research by Monbu-Kagakusyo of Japan (No. 21740182),
the U.S. Department of Energy under contract DE-AC02-98CH10886.
Numerical simulations have been performed on the BlueGene/L at the 
New York Center for Computational Sciences (NYCCS) which is supported 
by the U.S. Department of Energy and by the State of New York, and 
JUGENE at the John von Neumann Supercomputer Center (NIC) at 
FZ-J\"ulich, Germany, as well as the QCDOC computer of USQCD.


\begin{thebibliography}{99}

\bibitem{RHIC}
  I.~Arsene {\it et al.}, 
  Nucl.\ Phys.\ A {\bf 757}, 1 (2005)
  [arXiv:nucl-ex/0410020];
  B.~B.~Back {\it et al.},
  ibid. {\bf 757}, 28 (2005)
  [arXiv:nucl-ex/0410022];
  J.~Adams {\it et al.},  
  ibid. {\bf 757}, 102 (2005)
  [arXiv:nucl-ex/0501009];
  K.~Adcox {\it et al.},  
  ibid. {\bf 757}, 184 (2005)
  [arXiv:nucl-ex/0410003].

\bibitem{LeBellac}
  M.~Le~Bellac, {\it Thermal Field Theory}
  (Cambridge University Press, Cambridge, England 1996).

\bibitem{Fries:2003kq}
  R.~J.~Fries, B.~Muller, C.~Nonaka and S.~A.~Bass,
  Phys.\ Rev.\  C {\bf 68}, 044902 (2003)
  [arXiv:nucl-th/0306027].

\bibitem{fluctuations}
  R.~V.~Gavai and S.~Gupta,
  Phys.\ Rev.\  D {\bf 73}, 014004 (2006)
  [arXiv:hep-lat/0510044];
  S.~Ejiri, F.~Karsch and K.~Redlich,
  Phys.\ Lett.\  B {\bf 633}, 275 (2006)
  [arXiv:hep-ph/0509051].

\bibitem{newfluct}
  M.~Cheng {\it et al.},
  Phys.\ Rev.\  D {\bf 79}, 074505 (2009);
  C.~Schmidt,
  Prog.\ Theor.\ Phys.\ Suppl.\  {\bf 186}, 563 (2010)
  [arXiv:1007.5164 [hep-lat]].

\bibitem{plasmino}
  V.V.~Klimov, Sov. J. Nucl.\ Phys. {\bf 33}, 934  (1981)
  [Yad. Fiz. 33, 1734 (1981)];
  H.A.~Weldon, Phys. Rev. D {\bf 28}, 2007 (1983).

\bibitem{BBS92}
  G.~Baym, J.~P.~Blaizot and B.~Svetitsky,
  Phys.\ Rev.\ D {\bf 46}, 4043 (1992).

\bibitem{KKN06}
  M.~Kitazawa, T.~Kunihiro and Y.~Nemoto,
  Phys.\ Lett.\ B {\bf 633}, 269 (2006)
  [arXiv:hep-ph/0510167];
  Prog.\ Theor.\ Phys.\  {\bf 117}, 103 (2007)
  [arXiv:hep-ph/0609164].

\bibitem{Kitazawa:2007ep}
  M.~Kitazawa, T.~Kunihiro, K.~Mitsutani and Y.~Nemoto,
  Phys.\ Rev.\  D {\bf 77}, 045034 (2008)
  [arXiv:0710.5809 [hep-ph]].

\bibitem{Harada:2008vk}
  M.~Harada and Y.~Nemoto,
  Phys.\ Rev.\  D {\bf 78}, 014004 (2008)
  [arXiv:0803.3257 [hep-ph]].

\bibitem{Muller:2010am} 
  D.~Muller, M.~Buballa and J.~Wambach,
  Phys.\ Rev.\ D {\bf 81}, 094022 (2010)
  [arXiv:1002.4252 [hep-ph]].

\bibitem{Mueller:2010ah} 
  J.~A.~Mueller, C.~S.~Fischer and D.~Nickel,
  Eur.\ Phys.\ J.\ C {\bf 70}, 1037 (2010)
  [arXiv:1009.3762 [hep-ph]].

\bibitem{Qin:2010pc} 
  S.~-x.~Qin, L.~Chang, Y.~-x.~Liu and C.~D.~Roberts,
  Phys.\ Rev.\ D {\bf 84}, 014017 (2011)
  [arXiv:1010.4231 [nucl-th]].

\bibitem{Satow:2010ia}
  D.~Satow, Y.~Hidaka and T.~Kunihiro,
  Phys.\ Rev.\  D {\bf 83}, 045017 (2011)
  [arXiv:1011.6452 [hep-ph]].

\bibitem{Nakkagawa:2011ci} 
  H.~Nakkagawa, H.~Yokota and K.~Yoshida,
  Phys.\ Rev.\ D {\bf 85}, 031902 (2012)
  [arXiv:1111.0117 [hep-ph]].

\bibitem{Hidaka:2011rz} 
  Y.~Hidaka, D.~Satow and T.~Kunihiro,
  Nucl.\ Phys.\ A {\bf 876}, 93 (2012)
  [arXiv:1111.5015 [hep-ph]].

\bibitem{KK07}
  F.~Karsch and M.~Kitazawa,
  Phys.\ Lett.\  B {\bf 658}, 45 (2007)
  [arXiv:0708.0299 [hep-lat]].

\bibitem{KK09}
  F.~Karsch and M.~Kitazawa,
  Phys.\ Rev.\  D {\bf 80}, 056001 (2009)
  [arXiv:0906.3941 [hep-lat]].

\bibitem{Hamada:2010zz}
  M.~Hamada, H.~Kouno, A.~Nakamura, T.~Saito and M.~Yahiro,
  Phys.\ Rev.\  D {\bf 81}, 094506 (2010).

\bibitem{HTL}
  R. D. Pisarski, Phys. Rev. Lett. {\bf 63}, 1129 (1989);
  E.~Braaten and R.~D.~Pisarski,
  Nucl.\ Phys.\ B {\bf 337}, 569 (1990); ibid., {\bf B339}, 310 (1990).

\bibitem{KK09B}
  See, Appendix~B in Ref.~\cite{KK09}.

\bibitem{'tHooft:1976fv}
  G.~'t Hooft,
  Phys.\ Rev.\  D {\bf 14}, 3432 (1976)
  [Erratum-ibid.\  D {\bf 18}, 2199 (1978)].

\bibitem{Gattringer:2002mr}
  C.~Gattringer, R.~Hoffmann, S.~Schaefer,
  Phys.\ Lett.\  {\bf B535}, 358-362 (2002).
  [hep-lat/0203013].

\bibitem{Schafer:1996wv}
  T.~Schafer, E.~V.~Shuryak,
  Rev.\ Mod.\ Phys.\  {\bf 70}, 323-426 (1998).
  [hep-ph/9610451].

\bibitem{prep}
  O.~Kaczmarek, F.~Karsch, M.~Kitazawa, W.~S\"oldner,
  in progress.

\bibitem{distortion}
  H.~T.~Ding, A.~Francis, O.~Kaczmarek, F.~Karsch, E.~Laermann and W.~Soeldner,
  Phys.\ Rev.\  D {\bf 83}, 034504 (2011)
  [arXiv:1012.4963 [hep-lat]].
  {\bf [ADD MORE?] }

\bibitem{Peshier:1999dt}
  A.~Peshier and M.~H.~Thoma,
  Phys.\ Rev.\ Lett.\  {\bf 84}, 841 (2000)
  [arXiv:hep-ph/9907268].

\bibitem{Aarts:2002cc}
  G.~Aarts and J.~M.~Martinez Resco,
  JHEP {\bf 0204}, 053 (2002)
  [arXiv:hep-ph/0203177].

\bibitem{Petreczky:2005nh}
  P.~Petreczky and D.~Teaney,
  Phys.\ Rev.\  D {\bf 73}, 014508 (2006)
  [arXiv:hep-ph/0507318].

\bibitem{Meyer:2009jp}
  H.~B.~Meyer,
  Nucl.\ Phys.\  A {\bf 830}, 641C (2009)
  [arXiv:0907.4095 [hep-lat]].


\end{thebibliography}
\end{document}